**Research Article**

# Effect of Different Concentration and Application Method of Zinc on Yield of Chickpea (*Pisum Sativum L.*)

**Samaneh Goodarzi\*, Ali Tavanmand and Hamed Shajari**

*Department of plant physiology and Environmental science, Qom Azad University, Iran*

**\*Corresponding author:** Samaneh Goodarzi, Department of plant physiology and Environmental science, Qom Azad University, Iran.

## Abstract

This study evaluated the impact of Zinc (Zn) supplementation on the growth, yield, and seed quality of chickpea (Pisum sativum L.) under the semi-arid conditions of Kerman, Iran, across two growing seasons (2021-2022). A randomized complete block design was used with six treatments, including varying concentrations of zinc sulphate ($ZnSO_4$) applied via foliar spraying (0.1%, 0.25%, 0.5%) or irrigation (4, 8, 16 kg/ha), each applied twice-one and two months after greening. Results from the first year revealed significant differences in yield and yield components across treatments. The highest seed yield and protein content were achieved with foliar application of 0.5% $ZnSO_4$ and irrigation application of 8 kg/ha $ZnSO_4$. Zinc application enhanced reproductive processes, including pollen viability and stigma receptivity, leading to improved pod and seed attributes. However, excessive Zn application (e.g., 0.4% Zn) resulted in reduced plant performance, likely due to phytotoxicity.

Leaf Zn concentration was significantly higher with 16 kg/ha $ZnSO_4$ applied via irrigation, while foliar applications at 0.1% $ZnSO_4$ also increased Zn uptake efficiently. The second growing season, however, showed no significant differences in traits across treatments, which was attributed to favorable climatic conditions mitigating Zn deficiency. Zn deficiency remains a critical challenge globally, particularly in calcareous and nutrient-depleted soils, adversely affecting plant metabolism, root development, and nitrogen pathways. This study underscores the importance of optimizing Zn supplementation strategies to enhance yield and quality while avoiding toxicity. Findings provide practical recommendations for addressing Zn deficiencies in semi-arid cropping systems, offering valuable insights for sustainable chickpea production.

## Introduction

Chickpea (*Pisum sativum* L.) belongs to the Leguminosae family and holds significant importance as a global food legume, appreciated not only for its ancient domestication but also for its diverse roles as a vegetable, protein-rich pulse, and animal fodder. It ranks as the second most cultivated cool-season legume worldwide suited to temperate climates, with its origins tracing back around 9,000 years. Believed to have first emerged in the Middle Eastern regions of Syria, Iraq, and Iran, it has since been cultivated extensively in Europe and North America for centuries [1]. Canada stands as the world's leading producer of peas, followed closely by China, Russia, and India [2]. The application of zinc at appropriate concentrations and methods is crucial for soil health, as it not only enhances chickpea yield but also contributes to soil fertility by promoting nutrient cycling and microbial activity, similar to the benefits provided by legumes such as alfalfa or cover crops like rye, which enrich soil nutrients and improve soil structure [3-5]. It contains a variety of components including Carbohydrates, Starch, Dietary Fiber, Pro-teins (18 to 30%), Lipids, essential amino acids, Minerals and Vitamins, Polyphenols, β-carotene, zeaxanthin [6,3,7,].

Zinc is an essential micronutrient that acts as a crucial factor for proper function of numerous enzymes which involve in different processes such as nitrogen metabolism, protein production and energy movement. It is also integral to metabolism and various physiological processes, including oxidative reactions [8]. Zinc deficiency in crops not only limits yield but also affects quality by impairing metabolic processes. Zinc (Zn) is essential for nitrogen metabolism in plants, and its deficiency interferes with the production of proteins. Insufficient Zn availability negatively impacts root architecture, impairing root growth and development. As a result, plants experience reduced uptake of water and nutrients, which adversely affects their overall growth and yield. Moreover, reported that extreme Zn deficiency significantly limits the processes of flowering and fruit production, causing a notable decline in plant performance [9,10]. On the other hand, Excessive Zn concentrations in plants can lead to phytotoxicity, adversely affecting growth and yield. When leaf Zn levels reach approximately 300-1000μg Zn $g^{-1}$, yield reductions are commonly observed. The critical concentration for Zn-induced phytotoxicity is typically around 500μg Zn $g^{-1}$ [11].

The critical Zn thresholds in plants, as proposed by [12], are categorized as follows: less than 10 mg $kg^{-1}$ indicates a definite Zn deficiency, 10-15 mg $kg^{-1}$ suggests a very likely deficiency, 15-20 mg $kg^{-1}$ indicates a likely deficiency, and levels above 20 mg $kg^{-1}$ are considered sufficient. Crop and variety needs differ based on soil characteristics and climatic conditions. Nevertheless, a lack of Zn can result in lower yields or diminished product quality. For most crop species, the Zn sufficiency range in the dry matter of mature leaves is 15 to 50ppm, with 15 ppm generally regarded as the critical value for Zn adequacy [13]. Although Zn is commonly applied as a fertilizer, developing effective and cost-efficient strategies to address its deficiency in a sustainable manner tailored to specific cropping systems remains a priority. Proteins for human consumption can be sourced from both animal and plant origins. While the demand for animal-based proteins remains consistently high, they are often regarded as less environmentally sustainable [14].

In fact, reducing our use of animal sources will not only help us avoid possible harmful causes such as cardiovascular diseases [15], but also provide us a huge advantage to improve our eating habits in a more sustainable way. Peas, rich in essential nutrients and bioactive compounds, are an excellent addition to the human diet and hold great potential for a sustainable future. Therefore, developing effective strategies to enhance both the yield and quality of this valuable crop is of utmost importance. The study aimed to evaluate the impact of zinc applied through irrigation and foliar spraying, using zinc sulphate ($ZnSO_4$), on the productivity of chicken pea under the environmental condition of Kerman.

## Material and Methods

### Experimental Design and Location

The experiment was conducted over two growing seasons (2021-2022) at the Agricultural Research institution, kerman, Iran (30°55′57″N, 50°39′18″E), in order to evaluate the effect of different treatments of Zinc (Zn), both in concentration and method on the yield and yield components of Chicken Pea seeds. The site's semi-arid climate features cold winters, cool summers, and an average annual rainfall of 280mm, located 2,300 meters above sea level. The average annual temperature shows 10 degrees Celsius and the average annual rainfall is 318 millimeters.

The experimental design followed a randomized complete block format with 6 treatments which are **Treatment 1:** Control (no Zn application), **Treatment 2:** 4 kilograms of ZnSO4 per hectare as irrigation fertilizer in two stages, one and two months after greening, **Treatment 3**: 8 kilograms of ZnSO4 per hectare as irrigation fertilizer in two stages, one and two months after greening, **Treatment 4:** 16 kilograms of ZnSO4 per hectare as irrigation fertilizer in two stages, one and two months after greening **Treatment 5:** Foliar spraying with 0.25% concentration of ZnSO4 in two stages, one and two months after greening, **Treatment 6:** Foliar spraying with 0.5% concentration of ZnSO4 in two stages, one and two months after greening, **Treatment 7:** Foliar spraying with 0.1% concentration of ZnSO4 in two stages, one and two months after greening, **Treatment 8:** Foliar spraying with 0.4% concentration of ZnSO4 in two stages, one and two months after greenin

Each treatment plot with three replications measured 4 meters in length with a row spacing of 60 cm and plant spacing of 12 cm within rows. The soil was characterized as silty-loam, with a low organic carbon content (0.480%) and a boron concentration of 0.2mg/kg, making it an ideal test site for evaluating the impact of zinc supplementation. The experimental location also featured high pH levels (7.4), further challenging zinc availability to plants. These soil and climatic conditions mimic real-world challenges faced by farmers in semi-arid regions. For all treatments, a basal dose of 20kg/ha nitrogen (urea) was applied to ensure adequate nitrogen availability, also $ZnSO_4$ as source of Zn-fertilizer accompanied with essential nutrients were added to the soil (Table1). Foliar spraying included a surfactant was done to amplify absorption. For irrigation treatments, zinc sulfate was dissolved in water and evenly distributed across the plots using a drip irrigation system to ensure uniform application (Table 1).

**Table 1:** Some selected Physiological and chemical properties of the experimental field.

| Soil Parameter | Value |
| --- | --- |
| PH | 7.4 |
| Electrical conductivity | 2.36 |
| Organic carbon (%) | 0.48 |
| Available N (%) | 56 |
| Available P (mg/Kg) | 10.3 |
| Available K (mg/Kg) | 273.2 |
| Sand (%) | 42.2 |
| Loam (%) | 12.5 |
| Silt (%) | 40.3 |

**Data Collection and Analysis**

Soil samples were collected before the start of the experiment to establish baseline properties. Key agronomic parameters, such as seed yield, hundred-seed weight, zinc concentration in leaves, and protein content, were measured at harvest. Zinc concentrations in plant tissues were analyzed using atomic absorption spectrophotometry, while nitrogen and protein contents were determined using the Kjeldahl method. Statistical analysis was conducted using ANOVA, and treatment means were compared using Duncan's multiple range test at a significance level of 5%.

## Results and Discussion

The analysis of variance for the first year of the experiment revealed significant differences in the mean values of all traits. The findings demonstrated a notable variation in seed yield among treatments, significant at the 5% level. The maximum seed yield was achieved through the foliar application of a 0.50% Zinc solution, applied twice, once at one month and again at two months after emergence. Moreover, the application of 8 kilograms per hectare of Zn, through irrigation fertilizer treatments applied twice at one month and two months after emergence resulted in the highest yield (Table 2).

**Table 2:** Comparing the means of the measured traits in the first year using the Duncan test.

| Treatment | Number of Seeds/Pod | Weight of the Seeds/ Pod (g) | Zn Concentration in Leaf (mg/kg) | PROTEIN (%) |
|---|---|---|---|---|
| T1 | 4.36 | 1.85 | 32 | 21.35 |
| T2 | 6.22 | 2.5 | 45 | 23.62 |
| T3 | 6.56 | 2.65 | 58 | 22.77 |
| T4 | 6.35 | 2.37 | 56 | 23.65 |
| T5 | 6.31 | 2.76 | 60 | 23.76 |
| T6 | 6.87 | 2.76 | 59 | 24.87 |
| T7 | 6.84 | 2.7 | 62 | 23.48 |
| T8 | 6.02 | 2.01 | 38 | 22.18 |

**Table 3:** Comparing the means of the measured traits in the second year using the Duncan test.

| Treatment | Number of Seeds/ pod | Weight of the Seeds/ pod (g) | Zn Concentration in Leaf (mg/kg) | PROTEIN (%) |
|---|---|---|---|---|
| T1 | 4.36 | 1.85 | 32 | 24.35 |
| T2 | 4.22 | 1.82 | 33 | 23.62 |
| T3 | 4.3 | 1.8 | 30 | 23.87 |
| T4 | 4.35 | 1.83 | 34 | 23.65 |
| T5 | 4.47 | 1.79 | 31 | 25.76 |
| T6 | 4.4 | 1.84 | 30 | 23.72 |
| T7 | 4.39 | 1.87 | 33 | 23.48 |
| T8 | 4.37 | 1.78 | 32 | 23.72 |

Chickpea, like other salinity-tolerant plants such as barley (*Hordeum vulgare*) and quinoa (*Chenopodium quinoa*), exhibits crucial mechanisms for coping with saline conditions, including ion regulation, osmotic adjustment, and selective ion transport, which are further enhanced by adequate zinc application. Zinc plays a vital role in maintaining membrane integrity, activating antioxidant enzymes, and supporting osmoprotectant synthesis, making it essential for chickpea resilience in marginal and salt-affected soils [16] (*Munns, et al.,* 2006; *Ruiz Carrasco, et al.,* 2011; *Sohrabi sedeh et al.,* 2022). The beneficial effect of zinc on crop reproductive yield is primarily due to its role in enhancing male and female gametogenesis, which leads to a greater number of flowers produced per plant. Also, Zinc application promotes the development of sporogenous tissue, leading to a higher production of pollen grains per anther. Moreover, zinc enhances pollen-stigma interactions by improving stigma receptivity, functionality, and pollen viability. These combined effects ensure proper pollen germination, normal development, and significant improvements in yield attributes such as the number, size, and weight of pods and seeds [17]. our findings align with the results reported by many researchers such as [18-21].

On the other hand, our results indicate that applying Zn concentrations at 0.4% resulted in a reduction in values. This could be due to the fact that zinc, as a micronutrient, can have adverse effects on plant growth and development when applied in quantities exceeding the plant's requirements, potentially causing toxicity symptoms. Furthermore, since zinc is applied through foliar spraying, it is absorbed quickly and more efficiently by plants, with minimal loss compared to soil application. This efficiency, however, also increases the risk of negative effects, as even a slight excess can impact the plants adversely. From a broader perspective, the treatment with 0.50% zinc application (T6) emerges as the optimal approach, aligning with the zinc foliar application percentage recommended by [22] also [23]. As a result, Understanding the specific nutritional demands of Pea cultivars and evaluating soil properties in detail are critical for devising effective Zinc management strategies to optimize cultivation practices. Also, the analysis of variance for leaf Zn concentration revealed significance at the 1% level which according to the Duncan test indicated that the highest leaf Zn concentration was achieved with the application of 8 kilograms of Zn per hectare as an irrigation fertilizer, applied in two separate treatments, one and two months after emergence. Additionally, among the foliar application treatments, the highest leaf Zn amount was recorded with the application of a 0.1% Zinc solution (Table 2).

According to Table 2 the maximum seed protein content was achieved in Treatment Three, involving the application of 8 kilograms of Zn per hectare in two stages, one month and two months after greening. Moreover, the highest seed protein content among the foliar spray treatments was observed in Treatment seven, where 0.5% Zn was sprayed in two stages, one month and two months after greening. However, this result was not significantly different from the other foliar spray treatments. [16] investigated effect of Sulphur and Zinc on increase of soil health of Blackgram. They concluded that it can be attributed to the positive effects of zinc and sulfur in enhancing nodulation, which promoted higher atmospheric nitrogen fixation and, in turn, improved growth characteristics. The similar results were obtained by [24]. The results from the second year of the experiment, as presented in Table 3, reveal no significant variation among treatments for any of the measured traits.

The absence of significant differences among treatments can likely be attributed to climatic conditions. Zinc (Zn) deficiency seems to be linked to cooler and wetter seasons, with soil temperature influencing the rate of Zn mineralization [25]. Additional factors contributing to Zinc (Zn) deficiency in plants include high light intensity and extended day lengths [26,27]. In addition to natural soil and environmental factors, human-induced soil management practices often contribute to Zinc (Zn) deficiency. Moreover, plants may experience Zn deficiency under adverse climatic conditions, such as drought or soil compaction [28]. It seems that, unlike the first year of the project, temperature conditions during the second year were favorable, reducing the prominence of Zinc's role in plant growth. As a result, no significant differences in performance were observed among the treatments. The Duncan test comparison of average leaf Zn concentrations (Table 3) revealed that the highest leaf Zinc level was achieved with the application of 16kg of Zn per hectare, applied as an irrigation fertilizer in two stages, one and two months after greening. Zinc (Zn) is a critical micronutrient in plant metabolism, playing a key role in activating enzymes such as hydrogenase and carbonic anhydrase, which are essential for energy production and carbon assimilation. It also facilitates the structural integrity of ribosomal subunits, ensuring efficient protein synthesis, and contributes to the formation of cytochromes, which are indispensable for electron transport and respiration processes in plants [29]. Chickpea, like other salinity-tolerant plants such as fava bean (*Vicia faba*), exhibits crucial mechanisms for coping with saline conditions, including ion regulation, osmotic adjustment, and selective ion transport. The application of zinc, particularly through non-chemical fertilizers such as zinc-enriched compost or biofertilizers, can further enhance these mechanisms by improving enzymatic activity, supporting osmoprotectant synthesis, and boosting stress tolerance. Studies have demonstrated the importance of zinc-solubilizing bacteria in enhancing zinc uptake and plant growth under stress conditions (*Khan et al.,* 2021), while zinc oxide nanoparticles and biofertilizers have been shown to improve salinity tolerance and productivity in legumes [30] (*Rajput, et al.,* 2024; *Munir, et al.,* 2024). These approaches underscore zinc's critical role in sustainable agriculture, especially for crops grown in marginal and salt-affected soils (*Munns, et al.,* 2006; *Kumari, et al.,* 2019). Zinc (Zn) activates a range of plant enzymes that play pivotal roles in various physiological processes, including carbohydrate metabolism, preserving the stability of cellular membranes, facilitating protein synthesis, regulating auxin production, and supporting pollen development. [31,32].

Zinc deficiency in plants leads to the manifestation of various abnormalities, which are evident through symptoms such as stunted growth, chlorosis, reduced leaf size, and spikelet sterility. Zinc (Zn) deficiency, as a micronutrient, can negatively impact the quality of harvested crops. It also increases plants' vulnerability to damage caused by intense light or high temperatures and heightens their susceptibility to fungal infections [31,26]. Zinc (Zn) deficiency is a global issue, affecting soils in various regions, and nearly all crops show improved growth and yield when zinc is applied. Zinc (Zn) deficiency is commonly observed in calcareous soils, sandy soils, peat soils, and soils with elevated levels of phosphorus and silicon [27]. Research has shown that Zinc (Zn) plays a crucial role in the nitrogen metabolism pathway of plants. Its deficiency disrupts this process, leading to a significant reduction in protein synthesis. Zinc deficiency has a profound impact on the root system, hindering root development and reducing the plant's ability to absorb water and nutrients from the soil. This, in turn, leads to stunted growth and decreased yield [9]. Severe Zinc (Zn) deficiency has been shown to significantly impair the flowering and fruiting processes, leading to substantial reductions in reproductive success [10]. Zinc (Zn) deficiency is a significant challenge to global food production. Identifying Zn-deficient regions and understanding the underlying causes are essential for developing effective strategies to address this issue. While zinc fertilizers are commonly used, there is a need for more efficient and cost-effective methods to correct Zn deficiencies sustainably, tailored to specific cropping systems.

## Conclusion

This study highlights the critical role of zinc supplementation in improving the growth, yield, and quality of chickpea (*Pisum sativum* L.) under the semi-arid conditions of Kerman, Iran. The findings demonstrate that zinc application, particularly through foliar spraying at 0.5% concentration or irrigation with 8kg/ha $ZnSO_4$, significantly enhances seed yield, protein content, and zinc uptake. These effects are attributed to zinc's role in critical physiological processes, including nitrogen metabolism, pollen viability, and reproductive success. However, excessive zinc application, as seen with higher concentrations, can lead to phytotoxicity, emphasizing the importance of precise nutrient management tailored to crop and soil conditions.

The study also underscores the influence of climatic variability on zinc efficacy, as the second growing season showed no significant differences among treatments due to favorable weather conditions. This highlights the need for location-specific and seasonally adaptive nutrient management strategies to optimize zinc application.

**Future Insights**

Moving forward, there is a need for more extensive research to explore the following areas:

**Zinc Dynamics in Soil and Plant Systems:** Investigate the long-term effects of repeated zinc applications on soil health, zinc bioavailability, and crop nutrient balance.

**Cultivar-Specific Responses:** Examine the differential responses of various chickpea cultivars to zinc application to develop tailored recommendations for farmers.

**Climate Resilience:** Study the interaction between zinc supplementation and climatic stressors, such as drought and heat, to identify strategies for improving crop resilience under changing climatic conditions.

**Sustainable Zinc Application Methods:** Develop cost-effective and environmentally sustainable zinc delivery systems, such as nano-fertilizers or slow-release formulations, to minimize nutrient losses and environmental impacts.

**Integration with Other Nutrients:** Explore the combined effects of zinc with other micronutrients, such as boron or iron, to maximize crop productivity and quality. By addressing these future directions, the agricultural community can develop more robust,

## Acknowledgments

None.

## Conflict of Interest

None.

## References

1. Saskatchewan Pulse Growers (2000) Unlocking the bioavailability of phosphorus and micronutrient through development of low phytate phosphorus Pea. www.sask.pulse.com/ produce research project.
2. Raghunathan R, Hoover R, Waduge R, Liu Q, Warkentin TD (2017) Impact of Molecular Structure on the Physicochemical Properties of Starches Isolated from Different Field Pea (Pisum sativum L.) Cultivars Grown in Saskatchewan, Canada. Food Chem 221: 1514-1521.
3. Chel GL, Scilingo AA, Tintore SG, Davila G, Anon MC (2007) Physicochemical and structural characterization of Lima bean (Phaseolus lunatus) globulins. LWT. Food Sci. Technol 40: 1537-1544.
4. Mirbakhsh Mandana, Seyedeh Sara Sohrabi Sedeh, Zahra Zahed (2023) The impact of Persian clover (Trifolium resupinatum L.) on soil health. Black Sea Journal of Agriculture 6(5): 564-570.
5. Wu DT, Li WX, Wan JJ Hu, YC Gan, RY Zou (2023) A Comprehensive Review of Pea (Pisum sativum L.): Chemical Composition, Processing, Health Benefits, and Food Applications. Foods 12: 2527.
6. Kaur M, Sandhu KS, Singh N (2007) Comparative study of the functional, thermal and pasting properties of flours from different field pea (Pisum sativum L.) and pigeon pea (Cajanus cajan L.) cultivars. Food Chemistry 104: 259-267.
7. Han H, Baik BK (2008) Antioxidant activity and phenolic content of lentils (Lens culinaris), chickpeas (Cicerarietinum L.), peas (Pisum sativum L.) and soybeans (Glycine max) and their quantitative changes during processing. International J Food Sci. Technol 43: 1971-1978.
8. Hafeez B, Khanif YM, Saleem M (2013) Role of Zinc in Plant Nutrition- A Review. American Journal of Experimental Agriculture 3(2): 374-391.
9. Fageria NK (2004) Dry matter yield and nutrient uptake by lowland rice at different growth stages. Journal of Plant Nutrition 27(6): 947-958.
10. Epstein, Bloom (2005) Mineral Nutrition of Plants: Principles and Perspectives. Sinauer Assoc.
11. Chaney (1993) Risks associated with use of sewage sludge in Agriculture. In Proc. 15th Federal Convention.vol.1. Australian water and wastewater Association, Queensland, Australia.
12. Dobermann A, Fairhurst T (2000) Rice: Nutritional Disorders and Nutrient Management. Potash and Phosphate Institute and Potash and Phosphate Institute of 12. References 145 Canada (PPI/PPIC) and International Rice Research Institute (IRRI), Singapore and Makati City, the Philippines.
13. Benton JJ (2003) Agronomic handbook; management of crops, soils and their fertility. CRC press LLC. USA.
14. Langyan S, Yadava P, Khan FN, Dar ZA, Singh R, et al. (2022) Sustaining Protein Nutrition through Plant-Based Foods. Front. Nutr 8: 772573.

15. Naghshi S, Sadeghi O, Willett WC, Esmaillzadeh A (2020) Dietary Intake of Total, Animal, and Plant Proteins and Risk of All Cause, Cardiovascular, and Cancer Mortality: Systematic Review and Dose-Response Meta-Analysis of Prospective Cohort Studies. BMJ 370: m2412.

16. Kumar A, Swaroop N, David A, Thomas T (2020) Effect of Different Level of Sulphur and Zinc on Soil Health and Yield of Blackgram (Vigna mungo L.) Var. Barkha. Int J Curr Microbiol App Sci 9(8): 1175-1181.

17. Pandey N, Gupta B (2012) Improving seed yield of black gram (Vigna mungo L. var. DPU-88 31) through foliar fertilization of zinc during the reproductive phase. J Plant Nutr 35(11): 1683-1692.

18. Nadergoli MS, Yarnia M, Khoei FR (2011) Effect of Zinc and Manganese and Their Application Method on Yield and Yield Components of Common Bean (Phaseolus vulgaris L. cv. Khomein). Middle-East J Sci Res 8(5): 859- 865.

19. Salehin F, Rahman S (2012) Effects of zinc and nitrogen fertilizer and their application method on yield and yield components of Phaseolus vulgaris L. Agril Sci 3(1): 9-13.

20. Ali EA, Mahmoud AM (2013) Effect of foliar spray by different salicylic acid and zinc concentrations on seed yield and yield components of mung bean in sandy soil. Asian J Crop Sci 5(1): 33-40.

21. Borah L, Saikia J (2021) Effect of foliar application of zinc on growth and yield of garden pea (Pisum sativum L.) in Assam condition. International Journal of Chemical Studies 9(2): 869-872.

22. Päivöke A (1983) The long-term effects of zinc on the growth and development, chlorophyll content and nitrogen fixation of the garden pea. Ann Bot Fenn 20(2): 205-213.

23. Stoyanova Z, Doncheva S (2002) The effect of zinc supply and succinate treatment on plant growth and mineral uptake in pea plant. Braz J Plant Physiol 14(2): 111-116.

24. Abdul Baser, Zanier Shat, Muhammad Naeem, FehanBakht, Hfan ZH (2008) Effect of phos pharos and farm yard manure an agronomic traits of chickpea (Aicer arietinum L) Sarhad Journal of Agriculture 24(4): 567-572.

25. Takkar PN, Walker C (1993) The Distribution and Correction of Zinc Deficiency. In A.D. Robson (ed). Zinc in Soils and Plants (pp. 51). London: Kluwar Academic publisher.

26. Marschner H, Cakmak I (1989) High light intensity enhances chlorosis and necrosis in the leaves of zinc, potassium and manganese deficient bean (Physeolus vulgaris L.) plants. Plant Physiology 134: 308-315.

27. Cakmak I (2000) Role of zinc in protecting plant cells from reactive oxygen species. New Phytol 146:185-205.

28. Alloway BJ. (2008) Micronutrients and crop production. In Micronutrient Deficiencies in Global Crop Production. pp. 1-39© Springer Science Business Media BV.

29. Tisdale SL, Nelson WL, Beaten JD (1984) Zinc In soil Fertility and Fertilizers. Fourthedition, Macmillan Publishing Company, New York 382-391.

30. Mirbakhsh Mandana, Seyedeh Sara Sohrabi Sedeh (2023) The role of mycorrhiza and humic acid on quantitative and qualitative traits of faba bean plant under different fertilizer regimes. Ilmu Pertanian (Agricultural Science) 8(3): 175-185.

31. Marschner H (1995) Mineral nutrition of higher plants (2nd ed.). London: Academic Press.

32. Mirbakhsh Mandana (2022) "Effect of short and long period of salinity stress on physiological responses and biochemical markers of Aloe vera L." Ilmu Pertanian (Agricultural Science) 7(3): 178-187.